\documentclass[aps,prb,twocolumn,showpacs,superscriptaddress]{revtex4}
\usepackage{graphicx}
\usepackage{hyperref}
\usepackage{natbib}
\usepackage{bm}
\begin{document}

\title{Many-spin effects and tunneling splittings in 
Mn$_{12}$ magnetic molecules}
\author{H. A. De Raedt and A. H. Hams}
\address{Institute for Theoretical Physics and Materials Science Centre,
University of Groningen, Nijenborgh 4, NL-9747 AG
Groningen, The Netherlands}
\author{V. V. Dobrovitski, M. Al-Saqer, M. I. Katsnelson, B. N. Harmon}
\address{Ames Laboratory, Iowa State University, Ames, IA 50011, USA}

\begin{abstract}
We calculate the tunneling splittings in a Mn$_{12}$ magnetic
molecule taking into account its internal many-spin structure.
We discuss the precision and reliability of these calculations
and show that restricting the basis (limiting the number of
excitations taken into account) may lead to significant error
(orders of magnitude) in the resulting tunneling splittings
for the lowest energy levels, so that an intuitive picture of
different decoupled energy scales does not hold in this case. 
Possible routes for further development of the many-spin model of
Mn$_{12}$ are discussed.
\end{abstract}

%Keywords: tunneling, splittings, numerical diagonalization, magnetic
%molecules\\
%Corresponding author: Viatcheslav V. Dobrovitski\\
%Ames Laboratory, Iowa State University, Ames, IA 50011, USA\\
%phone: +1-515-294-8666, \ \ \ fax: +1-515-294-0689\\
%e-mail: slava@axel.ameslab.gov}

\maketitle

\section{Introduction}

Molecular magnets \cite{genintro} have proven to be very suitable 
systems for the
study of mesoscopic tunneling effects in magnetic materials. A
number of impressive experimental results have been obtained
recently, such as thermally-assisted tunneling 
\cite{jumps,jumps1}, ground state - to - ground state tunneling 
\cite{fe8tun,sangregorio} and topological phase effects in spin 
tunneling \cite{fe8tun}. Among others, the molecular magnet 
$\rm Mn_{12}O_{12}(CH_3COO)_{16}(H_2O)_4$ (below referred to 
as Mn$_{12}$) has received special attention. The 
effect of resonant magnetization tunneling has been first
observed and studied in detailed experiments \cite{jumps,jumps1}
on Mn$_{12}$, and, at present, a substantial amount of reliable
experimental data has been collected. Quantitative analysis of 
these experiments is a challenging theoretical problem involving 
fundamental issues about tunneling phenomena in
mesoscopic magnetic systems. The basic prerequisite for solving
this problem is our ability to evaluate accurately and reliably
the energy splittings occuring as a result of tunneling between
two (quasi)degenerate levels \cite{vanhemmen}. At present,
carefully designed magnetic relaxation experiments at low and
ultralow temperatures (tens or hundreds of milliKelvins) can 
detect \cite{jumps,jumps1} the changes in relaxation time
caused by the splittings of order 10$^{-2}$--10$^{-4}$ K, and
even smaller \cite{fe8tun}, of order 10$^{-6}$--10$^{-7}$ K. The
relaxation time data obtained in these experiments give
information (although indirect) about the splitting values, so
that predictions of the theoretical models can be compared with
experimental results. 

Conventionally, the molecular magnet Mn$_{12}$ is considered as a
large single spin ${\cal S}=10$ with quasidegenerate levels
${\cal S}_z=+M$ and ${\cal S}_z=-M$ split because of tunneling.
However, the single-spin Hamiltonian is a phenomenological 
construct; in reality, this is a many-spin system, consisting of
12 manganese ions coupled by exchange interactions. 
Here, using Mn$_{12}$ as a well-studied example, we address the
problem of reliable many-spin calculation of the tunneling
splittings in molecular magnets. Such a calculation is a very
complicated task. For example, the Hilbert space of the spin
Hamiltonian describing a molecule of Mn$_{12}$ consists of 10$^8$
levels, while the smallest tunneling splittings in Mn$_{12}$ are
of order of 10$^{-10}$ Kelvin (as measured in Ref.\ 
\cite{barbara} for $m=\pm 10$). The brute-force direct calculation
of tiny tunneling splittings in this system, even for several 
low-lying states,
is beyond the capabilities of modern computers. The general
strategy to solve this problem is to truncate the full Hilbert
space thus reducing consideration to a much smaller number of
relevant energy levels. This idea, implemented in a rather
sophisticated way, forms a basis of several approaches for the
evaluation of tunneling phenomena, such as quantum Monte-Carlo
methods \cite{qmcmod}, stochastic diagonalization \cite{deraedt},
and instanton calculations \cite{spindyn}.

To our knowledge, all calculations of the tunneling splittings in
molecular magnets starting from realistic Hamiltonians have
employed truncation of the Hilbert space in a much more
straightforward, and much less justified manner. High-energy
basis states, assumed to be irrelevant, are being explicitely
excluded from consideration, and only the low-energy part of the
spectrum is being taken into account \cite{mn12spl}. In the
present paper, we calculate tunneling splittings using the 
many-spin model of Mn$_{12}$, examining the accuracy and reliability
of this straightforward scheme. We demonstrate that, because of
strong Dzyaloshinsky-Morya interactions present in Mn12, the
splitting values obtained in this way are unreliable. We also
consider the sensitivity of the calculated splitting values to
variation in the Hamiltonian parameters, and determine the
accuracy needed for reliable splittings calculation.

The paper is organized as follows. In Section \ref{manyspin}, we
discuss the 8-spin model of Mn$_{12}$ and the methods used to
calculate tunneling splittings based on this model. We also
consider the stability of the results with respect to possible
limitations of the model Hamiltonians.
In Section \ref{discussion} we consider the reasons for the
failure of the energy-based truncation scheme in the 
splittings calculations. Our
conclusions can be found in the Summary.

\section{\label{manyspin}
8-spin model of M\lowercase{n}$_{12}$ and 
calculations of the tunneling splittings}

The cluster Mn$_{12}$ consists of eight Mn$^{3+}$ ions 
having spin 2 and four Mn$^{4+}$ ions having 
spin 3/2, coupled by exchange interactions.
The total number of spin states in Mn$_{12}$ 
is 10$^8$, and a corresponding Hamiltonian matrix
is rather large to be treated by modern computers. 
To overcome this difficulty, we can employ the natural hierarchy
of interactions present in Mn$_{12}$. The antiferromagnetic
exchange interactions $J_1\simeq 220$ K
between Mn$^{3+}$ and Mn$^{4+}$ ions are 
significantly stronger than all 
the others \cite{gat}, so corresponding pairs of Mn$^{3+}$ and 
Mn$^{4+}$ ions can be considered as stiff dimers with the total 
spin $s=1/2$, thus giving rise to the 8-spin model of Mn$_{12}$.
The range of validity of the 8-spin model, and the corresponding 
8-spin Hamiltonian of Mn$_{12}$ have been
considered in Ref.\ \onlinecite{mn12kat}. After examination 
of different possible interactions, the 
following Hamiltonian has been proposed:
\begin{eqnarray}
\label{hamilton}
{\cal H}&=& -J\Bigl(\sum_i {\bf s}_i \Bigr)^2
  -J'\sum_{\langle k,l\rangle} {\bf s}_k {\bf S}_l
  -K_z\sum_{i=1}^4 \left(S_i^z\right)^2\\ \nonumber
  && + \sum_{\langle i,j\rangle} {\bf D}^{i,j} \cdot
     [{\bf s}_i\times {\bf S}_j].
\end{eqnarray}
Here, ${\bf S}_i$ and ${\bf s}_i$ are the spin operators for the
large spins $S=2$ and small dimer spins $s=1/2$, correspondingly
(the subscript $i$ indexes the spins). The first two terms
describe isotropic Heisenberg exchange between the spins.
The third term describes the single-ion easy-axis anisotropy of
large spins. The fourth term represents the antisymmetric
Dzyaloshinsky-Morya (DM) interactions in Mn$_{12}$, where 
${\bf D}^{i,j}$ is the Dzyaloshinsky-Morya vector describing the 
DM-interaction between $i$-th small spin and $j$-th large spin.
Existence of DM-interactions in Mn$_{12}$ has been suggested
in Ref.\ \cite{barbara1}, and their magnitude has been estimated
in Ref.\ \cite{mn12kat} based on the neutron scattering data 
\cite{hennion}.
The molecules of Mn$_{12}$ possess a fourfold 
rotational-reflection axis (symmetry $S_4$) imposing
restrictions on the DM-vectors ${\bf D}^{i,j}$, so that
Dzyaloshinsky-Morya interactions can be described by only three
parameters $D_x \equiv D_x^{1,8}$, $D_y\equiv D_y^{1,8}$, and
$D_z \equiv D_z^{1, 8}$.

It has been demonstrated \cite{mn12kat} that the above model
satisfactorily describes a rather wide range of experimental data,
such as the splitting of the neutron scattering peaks,
results of EPR measurements and the
temperature dependence of magnetic susceptibility.
Here, for calculations we use the parameter set A from
Ref.\ \onlinecite{mn12kat}:
\begin{eqnarray}
\label{seta}
&&\text{set {\bf A}:}\\
  \nonumber
&&J=0, \quad J'=105 \text{ K}, \quad K_z=5.69 \text{ K}\\
  \nonumber
&&D_x=25 \text{ K}, \quad D_y=0, \quad D_z=-1.2 \text{ K}.
\end{eqnarray}
which also gives a good description of the response of Mn$_{12}$
molecules to a transverse magnetic field (external field
applied perpendicular to the easy axis of the
molecule). However, this set of parameters should not be
considered as being accurately determined, since the amount of 
the experimental information available is not yet sufficient to 
achieve particularly reliable parameters.
In the Hamiltonian (\ref{hamilton}), only the fourth term,
representing the Dzyaloshinsky-Morya (DM) interactions, can
lead to tunneling \cite{dm}: the first three terms conserve the
$z$-projection of the total spin ${\cal S}_z$ and can not induce
tunneling between levels with different ${\cal S}_z$, while
the DM-term mixes levels with different ${\cal S}_z$. In what 
follows, we will label the energy levels by the value of 
${\cal S}_z$. Although it is not an exact quantum number,
we can formally consider the DM-interaction as a perturbation,
and use perturbation theory terminology.

The following values of the tunneling splittings corresponding to
the parameter set (\ref{seta}) have been obtained by the
diagonalization of the full Hamiltonian matrix (of the size
$10^4\times 10^4$) using quadruple precision arithmetics:
\begin{eqnarray}
\label{spls}
\Delta E (\pm 10) &=& 1.18\cdot 10^{-15} {\rm K},\quad
\Delta E (\pm 8) = 1.06\cdot 10^{-11} {\rm K},\\
  \nonumber
\Delta E (\pm 6) &=& 3.87\cdot 10^{-8} {\rm K},\quad
\Delta E (\pm 4) = 2.08\cdot 10^{-6} {\rm K},\\
  \nonumber
\Delta E (\pm 2) &=& 4.17\cdot 10^{-2} {\rm K}.
\end{eqnarray}
The splittings for odd values of ${\cal S}_z$ are not shown: they
constantly remain at the level of the numerical precision of the
calculations (of order of 10$^{-19}$ K) \cite{cray}. In 
Mn$_{12}$, these splittings should be zero 
since the fourfold symmetry of the molecule imposes certain
restrictions on the symmetry of the spin Hamiltonian and makes
some matrix elements vanish. In the single-spin model of 
Mn$_{12}$ this property of the spin Hamiltonian is introduced
explicitly, by retaining only those operators which possess the
required fourfold symmetry. In the many-spin simulations, we
obtain the same result independently.

The first question to pose concerns the accuracy of the level
splitting evaluation. Parameters of the Hamiltonian are
determined with some finite precision, and a small error (say, of
the order of several Kelvin) affects the level energy by an
amount of order of Kelvin, which is much larger than the very
small value of tunneling splitting (of order of 10$^{-12}$ K).
Does it deprive the calculational results of all meaning? To
answer this question, we note that the levels 
$|{\cal S}_z=+M\rangle$ and $|{\cal S}_z=-M\rangle$ 
are degenerate due to exact symmetry properties of the spin
Hamiltonian, and, in the absence of the DM-term, would be
degenerate at any value of parameters. Therefore, the tunneling
splittings $\Delta E_{+M, -M}$ are governed only by the 
strength of the interaction which breaks the symmetry, i.e. the
DM-interaction. If the parameters of the Hamiltonian 
are determined with reasonably small {\it relative\/} error,
and if the numerical calculation is done with sufficient 
precision, then the 
{\it relative\/} error of the level splittings will also be 
small. This conclusion is supported by our calculations:
a 10\% variation in the Hamiltonian parameters
leads to the variation in the splitting values at most
by a factor of ten, so that accurate determination of
the Hamiltonian parameters is necessary for reliable calculation 
of the tunneling splittings. If only a logarithmic accuracy in the 
splitting values is needed, then the 10\% uncertainty in
the Hamiltonian parameters is sufficient.  

However, there is another, much more important source of possible
error. The description of the Mn$_{12}$ molecule by the 8-spin 
model requires a full, high-precision diagonalization of the 
Hamiltonian matrix with dimensions $10^4\times 10^4$.
Solving this problem is rather time-consuming.
Matrices of that size can be processed
very effectively using Lanczos-type methods, but the application
of these methods to the tunneling splitting calculations
constitutes quite a difficult problem by itself.
A very large number of iterations
is needed to achieve the necessary precision and in addition
the precision is hard to control when
the level separation is very small, so that special techniques are
necessary.

Therefore it is natural first to explore another approach,
namely, to omit high-energy basis states, retaining only the 
low-lying part of the spectrum where basis levels have energies 
less than some threshold value E$_{\rm cut}$. This approach has 
been adopted extensively and in fact, we are not aware of any
calculations of tunnel-splitting of magnetic molecules done in a
different way: calculations based on both the single-spin and the
many-spin model \cite{mn12spl} have employed this method. In this
paper, we assess the validity of this energy-based truncation
approach by considering the dependence of the tunneling
splittings $\Delta E_{+M, -M}$ for different pairs of degenerate
levels $|{\cal S} _z=+M\rangle$ and $|{\cal S}_z=-M\rangle$ on
the number of lowest levels $N_{\rm low}$ actually used in
calculations (or, in other words, their dependence on the energy
threshold $E_{\rm cut}$).

A brief description of the basis states is in order. We first
consider the first two exchange terms in the Hamiltonian of 
Eq.\ (\ref{hamilton}) and diagonalize within the manifold
of all the 8-spin configurations yielding states with 
${\cal S}_z=0$; there are 1286 energy eigenvalues corresponding
to eigenvectors with ${\cal S}$ ranging from 0 to 10. The 
distribution of states is: (10,1), (9,7), (8,24), (7,56),
(6,104), (5,164), (4,220), (3,248), (2,232), (1,168), (0,62),
where the first number in parenthesis is the value of ${\cal S}$
and the second is the number of levels with this value 
of ${\cal S}$. With the $2{\cal S}+1$ degeneracies included,
there are exactly 10000 states. These are the basis states which
are then used to diagonalize the full Hamiltonian, including
anisotropy and DM terms.

The initial increase in the number of basis states considered,
$N_{\rm low}$, leads to an 
overall increase in $\Delta E_{+M, -M}$
accompanied by oscillations (see Fig.\ \ref{figspl}).
After $N_{\rm low}$ achieves the value of about 700,
the oscillations have become small
and $\Delta E_{+M, -M}$ versus $N_{\rm low}$
exhibits a plateau.
This saturation lead in Ref.\ \onlinecite{mn12spl}
to the conclusion that the resulting values give the
actual splittings with sufficient accuracy.
But this conclusion is wrong.
A further increase of the number of levels leads to a 
resurrection of the oscillations at $N_{\rm low}\sim 1200$, 
with a quite pronounced jump in $\Delta E_{+M, -M}$
for $N_{\rm low}\sim 1700$. For a larger number of
levels, the situation repeats itself: the 
values of the splittings reach another plateau,
then oscillations appear again with a subsequent jump, etc.
We have traced this behavior up to $N_{\rm low}\sim 3000$,
which is already 1/3 of the total number of levels.
The observed behavior of $\Delta E_{+M, -M}$
is, in our opinion, a very clear signal that
energy-based truncation of the Hilbert space
is not a good strategy for the computation of tunneling 
splittings: it gives unreliable results.

The rather sharp jumps in the tunneling splittings as discussed
above and illustrated in Fig.\ \ref{figspl} are associated with
the inclusion of basis states with large ${\cal S}$ values. Because
of the selection rule for the DM term 
(${\cal S}\to {\cal S}\pm 1$), the ${\cal S}=10$ ground state
only couples with ${\cal S}=9$ states. States with smaller 
${\cal S}$ values affect the splittings more indirectly by
coupling with other states which eventually couple to the ground
state. While the states with large ${\cal S}$ cause jumps in the
splitting values, there are few of them, and the smaller coupling
of smaller ${\cal S}$ states still is significant because of the
cumulative effect of so many states (see the distribution given
above). Therefore, the evaluation of tunneling splittings for
a general system possessing strong DM interactions requires
consideration of sufficiently large portion of Hilbert space.

It is noteworthy that
the same truncation method works rather well for 
calculations of the energies of well-separated levels.
To compare the model against most of the
experiments, it suffices to know the positions of the levels with
much less precision, usually an error less than
0.1 K is already adequate. This level of precision
can be obtained by taking into account
$N_{\rm low}\sim 1000$ levels (i.e., 1/10 of the total
Hilbert space). Even using $N_{\rm low}\sim 500$,
the error in the level position is less than
1 K even for the states of energy about 60 K.
Therefore, the matrix-truncation approach is
adequate for fitting the model parameters to experimental
data. But the calculations of the tunneling splittings 
should be done using the full Hamiltonian matrix.

\section{\label{discussion}Discussion}

We have shown that truncating the Hilbert
space leads to large errors in the calculated values
of tunneling splittings. But actually, any sensible Hamiltonian
is inevitably obtained due to some truncation
of the Hilbert space.
For example, the Hamiltonian (\ref{hamilton}) can be considered
as a result of the two-step procedure \cite{yosida}: (i)
projection of the real many-electron Hamiltonian onto the
subspace of suitably chosen single-electron orbital states,
yielding a general spin Hamiltonian of the molecule; and (ii)
projection of the resulting spin Hamiltonian onto the subspace of
the 8-spin model. This procedure is usually justified
(at least, at the heuristic level)
by invoking some kind of perturbation or WKB-theory arguments, 
and corresponds to an intuitive idea of different, practically
independent energy scales.

However, in the case of the tunneling splittings, we see that
very different energy scales significantly affect each other.
Why do the same arguments not
work if we truncate the 8-spin Hamiltonian?
In our opinion, this takes place because the conditions of 
the applicability of WKB-reasoning (or similar arguments
based on perturbation theory) are not satisfied.
% There are
% two possible causes for this. 
%
%First, t
The spin of the
system ${\cal S}=10$ is too small, so that the instanton
action \cite{vanhemmen} on the trajectories corresponding to the
8-spin model is not large enough. Indeed,
for systems with well-separated levels, the quasiclassical
approximation usually already works reasonably for a total
spin ${\cal S}\sim 2$--3. However, as has been
demonstrated \cite{levine}, to apply the same type
of arguments to the splitting calculations, the
(normalized) instanton action $S_I$ should exceed the value
of 12. 
For the model employed in Ref.\ \onlinecite{levine},
this corresponds to the system with a 
total spin (more exactly, with the total
antiferromagnetic vector) of order of several thousand.
Thus, the tunneling splittings, in general, appear to be 
much more sensitive to the method of calculation
than the level energies themselves, and conditions
for applicability of the conventional WKB-reasoning
are considerably more stringent (though
for Mn$_{12}$ they can of course be different from the condition 
$S_I>12$).
Qualitatively this agrees with our observations (see Section
\ref{manyspin}). Even
a rather severe truncation of the Hilbert space has a minor
effect on the level energies, 
while correct values of
the tunneling splittings require a diagonalization of the full
Hamiltonian.
% The second argument concerns the high
% symmetry of the Hamiltonian, which manifests itself
% in a large number of vanishing matrix elements
% (such as, for example, the zero splittings $\Delta E_{+M, -M}$
% for odd $M$). It leads to a well-pronounced
% non-smoothness of the splittings as a function of the
% number $N_{\rm low}$ of the levels taken into account
% (see Section \ref{manyspin}), preventing the use of a WKB 
% approach.

Briefly, these arguments can be expressed in a rather
obvious form: the 8-spin model is not ``macroscopic
enough'' to justify the truncation of the Hilbert space
by some WKB or similar perturbation approach. In this
case the intuitive picture of different independent energy
scales is misleading.

This conclusion raises important questions, namely, is the 8-spin
model, being the result of the truncation of, e.g., 12-spin
Hamiltonian, sufficient to predict reliably the tunneling
splittings (or, in other words, is the 12-spin model
``macroscopic enough'' to be truncated)? What is the minimal
model allowing the splittings to be calculated correctly? We 
believe that these are key questions, not only for Mn$_{12}$ 
but for the whole class of magnetic molecules. 
%
% In particular, the applicability of the single-spin model
% for quantitatively correct splittings calculations
% is questionable. Currently, the results obtained by using
% this model are not satisfactory
% (see Section \ref{singlespin}). 
%
For this purpose, ab-initio calculations of the
exchange and anisotropic intramolecular interactions in Mn$_{12}$
could be very useful. Also, reliable experimental data for the
tunneling splittings would obviously be of great value for
further development.

\section{Summary}

We have calculated the tunneling splittings in
Mn$_{12}$ on the basis of the 8-spin model proposed
earlier \cite{mn12kat}. We have shown that rather accurate
knowledge of the Hamiltonian parameters is needed for the 
accurate splitting calculations; although, for logarithmic
accuracy, 10\% error in the parameters can be tolerated.
Furthermore, we have demonstrated that a
reliable calculation of the tunneling splittings for
a system with strong DM interactions requires
the use of the full Hamiltonian matrix. We have explicitely
shown that an energy-based Hilbert 
space truncation scheme can be successfully used
for the determination of the level energies, but leads to
erroneous results when applied to the splitting calculations.

\section*{Acknowledgments}

This work was partially carried out at the Ames Laboratory, which
is operated for the U.\ S.\ Department of Energy by Iowa State
University under Contract No.\ W-7405-82 and was supported by the
Director of the Office of Science, Office of Basic Energy 
Research of the U.\ S.\ Department of Energy.
Support from the Dutch ``Stichting Nationale Computer
Faciliteiten (NCF)'' and the Dutch ``Stichting voor Fundamenteel
Onderzoek der Materie (FOM)'' is gratefully acknowledged.

\begin{figure}
\caption{Dependence of the tunneling splittings 
$\Delta E_{+M, -M}$ (in Kelvins) versus the number of levels 
taken into account in the many-spin calculations. The parameter
set A (see text) has been used for calculations. The results 
for $M=8$, 6, 4, and 2 are presented. Tunneling splittings for 
the levels with odd $M$ are zero because of the symmetry 
properties of the spin Hamiltonian.}
\label{figspl}
\end{figure}


\begin{thebibliography}{99}
\bibitem{dm} Note that Dzyaloshinsky-Morya interactions have 
  nonzero matrix elements only for the pairs of
  levels $|{\cal S},{\cal S}_z\rangle$ and 
  $|{\cal S}\pm 1,{\cal S}_z\pm 1\rangle$, but they do not
  couple the levels with the same value of the total spin
  $\cal S$. It means that the tunneling splittings are
  governed by the ratio $D_{x,y,z}/J'$, i.e. the 
  tunneling barrier is created primarily by the exchange.
  This is in contrast with the single-spin model picture,
  where the tunneling appears due to the anisotropy
  term ${\cal S}_+^4 + {\cal S}_-^4$, so that the ratio
  $\gamma /\alpha$ determines the splittings, and the
  tunneling barrier coincides with the anisotropy barrier.
\end{thebibliography}
\end{document}